\documentclass[twocolumn]{aastex631}

\DeclareUnicodeCharacter{2212}{-}
\usepackage{multirow}
\usepackage{comment}
\usepackage[normalem]{ulem}
\usepackage{amsmath}

\shorttitle{Lithium Signatures Post-Planet Engulfment}
\shortauthors{Sun et al.}

\graphicspath{{./}{figures/}}
\begin{document}
		
\title{C3PO III: On the Lithium Signatures Following Planet Engulfment by Stars}
		
\correspondingauthor{Qinghui Sun}
\email{qinghuisun@sjtu.edu.cn}
		
\author[0000-0003-3281-6461]{Qinghui Sun}
\affiliation{Tsung-Dao Lee Institute, Shanghai Jiao Tong University, Shanghai, 200240, China}
\affiliation{Department of Astronomy, Tsinghua University, Beijing, 100084, China}
		
\author[0000-0001-5082-9536]{Yuan-Sen Ting}
\affiliation{Department of Astronomy, The Ohio State University, 1251 Wescoe Hall Dr., Columbus, Ohio, 43210, USA}
\affiliation{Center for Cosmology and AstroParticle Physics, The Ohio State University, 191 West Woodruff Avenue, Columbus, Ohio, 43210, USA}
		
\author[0000-0003-4794-6074]{Fan Liu}
\affiliation{School of Physics and Astronomy, Monash University, Melbourne, VIC, 3800, Australia}
\affiliation{ARC Centre of Excellence for Astrophysics in Three Dimensions (ASTRO-3D), Canberra, ACT, 2611, Australia}
		
\author[0000-0002-6937-9034]{Sharon Xuesong Wang}
\affiliation{Department of Astronomy, Tsinghua University, Beijing, 100084, China}
		
\author[0000-0001-8841-3579]{Barbara J. Anthony-Twarog}	\author[0000-0001-5436-5206]{Bruce A. Twarog}
\affiliation{Department of Physics and Astronomy, University of Kansas, 1251 Wescoe Hall Dr., Lawrence, KS, 66045, USA}
		
\author[0000-0002-6332-0453]{Jia-Yi Yang}
\author[0000-0003-0707-3213]{Di-Chang Chen}
\affiliation{School of Astronomy and Space Science, Nanjing University, Nanjing, 210023, China}
		
\author[0000-0002-3625-6951]{Amanda I. Karakas}
\affiliation{School of Physics and Astronomy, Monash University, Melbourne, VIC, 3800, Australia}
\affiliation{ARC Centre of Excellence for Astrophysics in Three Dimensions (ASTRO-3D), Canberra, ACT, 2611, Australia}
		
\author[0000-0002-6472-5348]{Ji-Wei Xie}
\affiliation{School of Astronomy and Space Science, Nanjing University, Nanjing, 210023, China}
		
\author[0000-0002-6502-1406]{David Yong}
\affiliation{Research School of Astronomy \& Astrophysics, Australian National University, Cotter Rd., Weston, ACT, 2611, Australia}
		
\begin{abstract}
			
Planet engulfment has been identified as one of the mechanisms for enhancing lithium abundance in stars. However, comprehensive investigations into lithium signatures following such events remain limited. Stars born together, sharing a common origin and stellar characteristics, provide a unique opportunity to study these signatures and compare lithium abundances. We demonstrate that the distinctive signature of planet engulfment in lithium abundance is only discernible among highly similar stellar twins. We present lithium abundance measurements for 125 co-moving pairs of stars, representing the largest sample to date with a single, homogeneous assessment of high-precision lithium abundance. While lithium abundance enhancements in pairs showing planet engulfment signatures are within 0.35 dex, we find that even at fixed stellar parameters (temperature and age), the intrinsic scatter in lithium abundance is typically 0.35 dex for G/F dwarfs and can be as large as 0.6 dex for older and cooler stars due to internal stellar evolution processes. Since the planet engulfment signature from lithium can be masked by stellar intrinsic scatter, our findings raise questions about relying solely on lithium as an indicator for planet engulfment or attributing lithium-richness in stars primarily to planet engulfment events.

\end{abstract}
		
\section{Introduction}
		
Lithium (Li) was initially formed during Big Bang nucleosynthesis (\citealt{2000ApJ...530L..57R, 2014JCAP...10..050C, 2016RvMP...88a5004C, 2021A&A...653A..48D}), with additional production occurring in the galaxy (\citealt{2015Natur.518..381T, 2019A&A...623A..99G}). Within stars, lithium is easily depleted in the hot stellar interior (\citealt{1997ARA&A..35..557P, 2017AJ....153..128C, 2023ApJ...952...71S}), but it is believed to be produced through both internal and external processes. These include the Be$^7$ transport mechanism (\citealt{1971ApJ...164..111C, 2015ApJ...808L..14I, 2022MNRAS.513.5387S}) and the engulfment of Li-rich substellar companions such as brown dwarfs or planets (\citealt{1999MNRAS.308.1133S, 2016ApJ...829..127A, 2018ApJ...854..138O, 2021ApJ...922..129G, 2023A&A...670A.155C}). However, the extent and duration of changes in A(Li) resulting from planet engulfment are complex, contingent upon factors such as composition, mass, and timing of the engulfment process (\citealt{2022MNRAS.516.3354S, 2023MNRAS.518.5465B}). These changes are further influenced by subsequent internal processes including convection, diffusion, and thermohaline mixing (\citealt{2010A&A...521A...9C, 2015MNRAS.446.2673L, 2022MNRAS.516.3354S}). For instance, \cite{2021AJ....162..273S} found that engulfment of a planet is not a feasible enrichment mechanism for stars that have evolved beyond the subgiant phase.
		
Furthermore, while planet engulfment can produce observable alterations in lithium abundance, the intrinsic scatter of Li remains a major contributing factor, which could potentially mask the signal from planet engulfment (e.g., \citealt{2022GeAe..62..903K, 2020MNRAS.492..245C, 2018AA...613A..63B, 2014AA...562A.102O, 1997AA...323...86R, 1990ApJ...354..310B}). Despite this known complication, there is still a lack of observational studies with consistent stellar parameter measurements and a quantitative analysis of lithium signatures following planet engulfment by main-sequence stars.
		
Co-natal pairs of stars, sharing a common origin and similar stellar characteristics, offer a unique advantage for identifying planet engulfment signatures and directly comparing A(Li) within the pair. Multiple studies have identified planet engulfment signatures within co-natal pairs by analyzing the condensation temperature trend of various elements (e.g., \citealt{2017A&A...604L...4S, 2020ApJ...888L...9N}). However, only a few studies have rerpoted lithium in the context of comoving stars (e.g., \citealt{2019A&A...628A.126M, 2018ApJ...854..138O}). Futhermore, often inhomogeneities in stellar parameter measurements between pairs of co-natal stars make comparisons across studies challenging. Due to the lack of a large sample of co-natal pairs with uniform stellar parameter measurements and the complexity of $\Delta$A(Li), studying the differences between pairs that show post-planet engulfment and those that do not remains difficult. Consequently, changes in A(Li) following planet engulfment are still poorly understood.
		
In this study, we aim to address this gap. We report A(Li) and associated uncertaintys for 125 pairs of stars (250 stars total), including 89 co-natal pairs, of which 7 pairs have reported planet engulfments. These stars were observed using 6-10m class telescopes with high spectral resolution and high signal-to-noise ratios. Such exquisite spectra allow us to consistently measure high-precision A(Li) ($\sigma$ $\sim$ 0.01 to 0.04 dex) to study the extent of lithium enhancement following planet engulfment and compare it to theoretical models.
				
\section{Data Selection} 
		
This paper builds upon the previous two C3PO (Complete Census of Co-moving Pairs Of stars) papers (\citealt{2023MNRAS.526.2181Y}, \citealt{Liu2024}). High-resolution (R = $\lambda$/$\Delta\lambda$ $\sim$ 50,000-110,000), high signal-to-noise ratio (S/N $\approx$ 250 pixel$^{-1}$) spectra have been obtained for 125 co-moving pairs of stars (250 stars total) using the Magellan Telescope, the Keck Telescope, and the Very Large Telescope. Notably, C3PO selects comoving stars that appear to be stellar twins (similar colors and magnitudes, and hence similar effective temperature and surface gravity). C3PO I (\citealt{2023MNRAS.526.2181Y}) derived stellar parameters for these stars and defined two groups: chemically homogeneous pairs and chemically anomalous pairs ($|\Delta$[Fe/H]$|$/$\sigma$[Fe/H] $>$ 3.0), where $\Delta$ represents the difference between the pair of stars, and $\sigma$ represents the statistical uncertainty of the measurements. The effective temperature ($T_{\rm eff}$) of the stars ranges from 4900 K to 6500 K. In C3PO I, high-precision stellar parameters were derived, with average uncertainties of 20 K in $T_{\rm eff}$, 0.03 in surface gravity (log g), and 0.014 dex in metallicity ([Fe/H]). Table 1 displays detailed parameters for the comoving pairs.
		
\begin{deluxetable*}{cccccccccccc}
			\label{Table1}
			\tablecaption{Comoving pairs from C3PO and their corresponding properties adopted this study}
			\tabletypesize{\scriptsize}
			\tablewidth{1.0\textwidth}
			\setlength{\tabcolsep}{2pt}
			\decimalcolnumbers
			\renewcommand{\arraystretch}{1.2}
			\tablehead{
				pair ID & ${\rm star}_{\rm ref}$ & Gaia DR3 ID & $T_{\rm eff, ref}$ & $\log g_{\rm ref}$ & ${\rm [Fe/H]}_{\rm ref}$ & $V_{\rm t,ref}$ & $V_{\rm ROT,ref}$ & uncertainty & ${\rm age}_1$ & ${\rm uncertainty}_1$ & ... 
			}
			\startdata
			\multicolumn{12}{c}{{\bf Co-natal Pairs$^6$, homogeneous sample} with $|\Delta T_{\rm eff}|$ $<$ 100 K, $|\Delta$log g$|<$ 0.1, $T_{\rm eff}$ $>$ 6000 K (17 pairs)} \\
			3 & sk183a & 692119656035933568 & 6003 & 4.56 & -0.355 & 1.04 & 20.50 & 1.17 & 3.67 & 0.45 & ... \\
			... & ... & ... & ... & ... & ... & ... & ... & ... & ... & ... & ... \\[0.2cm]
			\hline
			\\[-0.3cm]
			\multicolumn{12}{c}{{\bf Other co-natal pairs}$^6$ (72 pairs)} \\
			1 &  star29a  & 55780840513067392    & 5986  & 3.97  & 0.082 & 1.43  & 5.25  & 0.02  &     3.60    & 0.02  & ... \\
			... & ... & ... & ... & ... & ... & ... & ... & ... & ... & ... & ... \\[0.2cm]
			\hline
			\\[-0.3cm]
			\multicolumn{12}{c}{{\bf Non co-natal pairs}$^6$, with log$_{10}$($\Delta$s) $>$ 6 or  log$_{10}$($\Delta$v) $>$ 2 (36 pairs)} \\
			6 & sk219a &  775037328283498624   & 5836  & 4.47  & -0.027    & 1.37  & 17.48 & 0.86  & 5.08  & 0.60   & ... \\
			... & ... & ... & ... & ... & ... & ... & ... & ... & ... & ... & ... \\[0.2cm]       
			\hline
			\enddata\tablecomments{1. The Pair ID and star reference name (columns 1 and 2) follow the C3PO designation as in \citet{2023MNRAS.526.2181Y} and \citet{Liu2024}. 
				2. Columns 4-17 present basic stellar parameters for the reference star: effective temperature ($T_{\rm eff}$), surface gravity (log g), metallicity ([Fe/H]), and microturbulence ($V_t$). These parameters are adopted from C3PO I \citep{2023MNRAS.526.2181Y}. Ages are computed from Gaia parallax (age1) and spectroscopic log g (age2) using the q$^2$ package and $Y^2$ isochrones. Rotational velocity ($V_{\rm ROT}$) and its uncertainty are reported by {\it fxcor}. We also report the signal-to-noise ratio (S/N) near the Li line, 1D Li abundance (A(Li)), 3D non-LTE corrected A(Li), and the combined uncertaintys from 1$\sigma$ EW measurement and stellar atmosphere propagation.
				3. Columns 18-33 display the same parameters as columns 2-17, but for the object star. We note that the reference and object stars are randomly assigned within each pair.
				4. Columns 34-37 are the uncertaintys in $T_{\rm eff}$, log g, [Fe/H], and $V_t$.
				5. Columns 38-40 are the logarithmic values of the pair's spatial separation ($\Delta$s) in AU and velocity dispersion ($\Delta$v) in km s$^{-1}$, respectively, along with the difference in Bayesian evidence ($\Delta$ln(Z)) between the planet engulfment and null hypothesis (flat differences in abundances) as described in \citet{Liu2024}. A higher $\Delta$ln(Z) favors the star having undergone planet engulfment.
				6. We separate co-natal pairs ($\Delta$s $<$ 10$^6$ AU and $\Delta$v $<$ 2 km s$^{-1}$) from non-co-natal pairs by their spatial and velocity offsets. Within the co-natal pairs, we identify a homogeneous sample with $|\Delta T_{\rm eff}|$ $<$ 100 K, $|\Delta$log g$|<$ 0.1, and $T_{\rm eff}$ $>$ 6000 K (17 pairs). Seven pairs of all the co-natal pairs (69, 74, 77, 79, 112, 116, and 124) meet all criteria for having engulfed planet(s) as described in \citet{Liu2024}, showing strong evidence of planet engulfment, and are marked with double asterisks next to their pair IDs. Four other pairs (15, 33, 62, and 122) meet the $\Delta$ln(Z) criterion but not all other criteria, and are marked with a single asterisk next to their pair IDs. \\
				(This table is available in its entirety in machine-readable form.)
			}
\end{deluxetable*}

C3PO I found that comoving stars with separations ($\Delta$s) closer than 10$^6$ AU appear to be substantially more chemically homogeneous than their more distant counterparts. Based on this, the study proposed a potential boundary for distinguishing between co-natal and non-co-natal pairs from these comoving stars, using both 3D spatial separation and velocity difference ($\Delta$v). In Table 1, we distinguish the 89 co-natal pairs that satisfy $\Delta$s $<$ 10$^6$ AU and $\Delta$v $<$ 2 km s$^{-1}$ from the non-co-natal pairs of stars. Within the co-natal pairs, the differential stellar parameters are constrained to $|\Delta T_{\rm eff}|$ $<$ 300 K, $|\Delta$log g$|<$ 0.3, and $|\Delta$[Fe/H]$|$ $<$ 0.3 dex, with average values of $|\Delta T_{\rm eff}|$ = 169 K, $|\Delta$log g$|$ = 0.07, and $|\Delta$[Fe/H]$|$ = 0.06 dex, demonstrating that most of these comoving stars are indeed stellar twins. For more detailed information, please refer to \cite{2023MNRAS.526.2181Y}.
		
C3PO II (\citealt{Liu2024}) further derived abundances for 21 elements (excluding Li) from these comoving stars. As most of these pairs are stellar twins, exquisite precision can be achieved for the difference between abundances through line-by-line differential analysis. \citealt{Liu2024} reported an uncertainty of $\sim$ 0.015 dex (3.5\%), covering elemental abundances spanning a range of condensation temperatures ($T_{\rm cond}$). Based on this, C3PO II established an analytical framework to detect planet engulfment signatures in stellar pairs. The framework involves comparing the difference in Bayesian evidence ($\Delta$ln(Z)) between the abundance patterns of the pairs of stars for planet engulfment scenario and the null case, and evaluating the elemental abundance ([X/H]) -- condensation temperature ($T_{\rm cond}$) trends. Out of the 89 co-natal pairs, 21 meet the $T_{\rm cond}$ trend criterion, 11 meet the $\Delta$ln(Z) criterion, and seven meet both criteria. 
		
In Table 1, the seven pairs identified by C3PO II are denoted with double asterisks next to their pair IDs, which are deemed to be genuine signals from planet engulfments. C3PO II also modelled and argued that effects of atomic diffusion are unlikely to explain the abundance trend. There are four probable pairs meeting the $\Delta$ln(Z) criterion, but not the temperature trend, which are marked with a single asterisk. From the seven pairs, they inferred that the average mass of accreted material for the planet-engulfed pairs is 4.3 $\pm$ 0.8 M$_{\oplus}$, with a standard deviation of 2.1 M$_{\oplus}$.
		
\begin{figure*}[h]
		\centering	\includegraphics[width=0.8\textwidth]{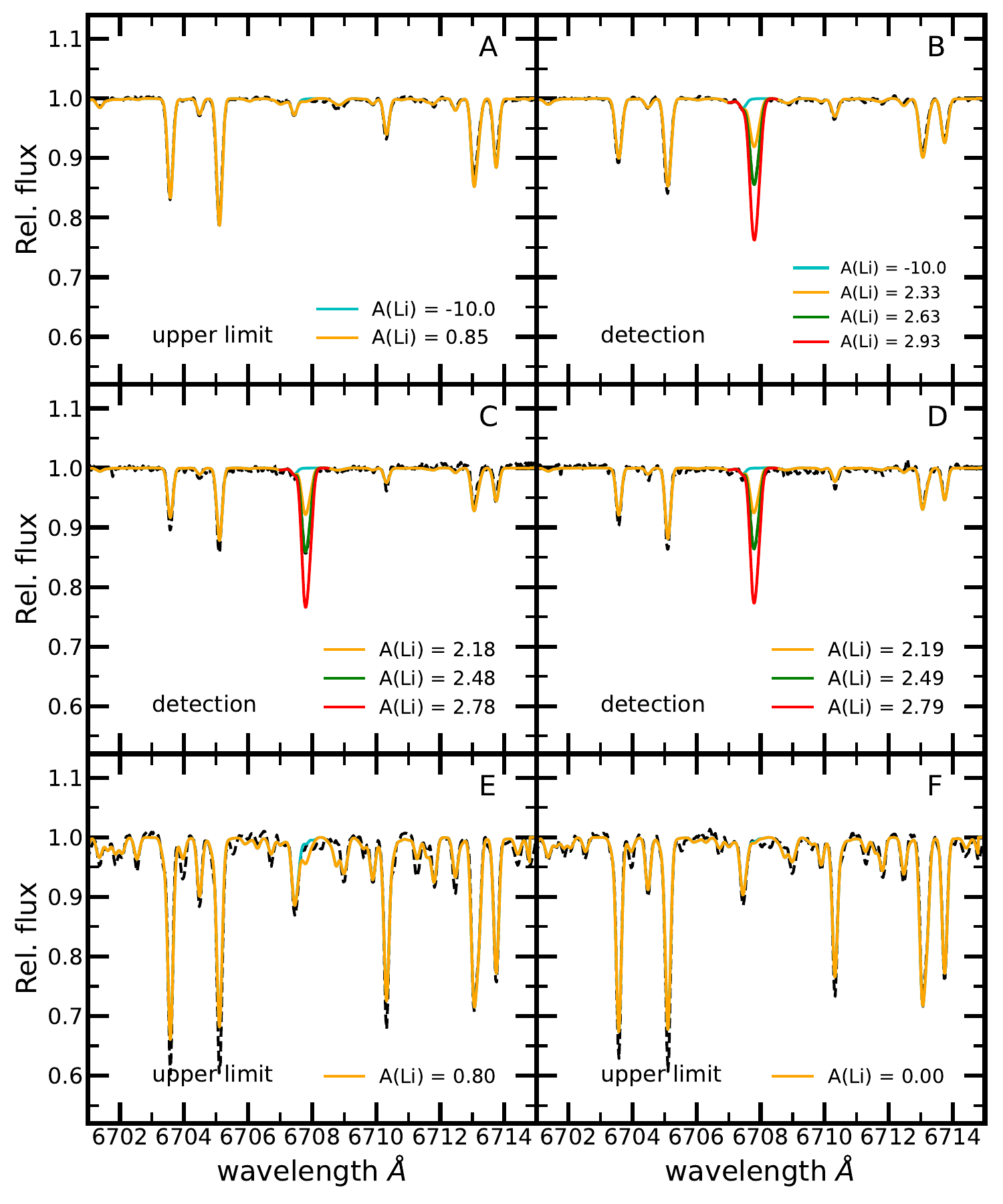}
		\caption{The observed (black dashed line) and synthetic spectra (solid colored lines) for three stellar pairs, showcasing both lithium detections and 3$\sigma$ upper limits. For Li detections, the green line displays the best-fit A(Li), while the orange and red lines show A(Li) values that are half and twice the best-fit A(Li), respectively. In cases where no confident Li detection is made, we show the 3$\sigma$ upper limit A(Li) (orange line) that corresponds to the computed 3$\sigma$ EW. The no Li case is always displayed as a cyan line. The Li detection case is notably distinguishable from the upper limits due to a much stronger Li absorption line at 6707.8\AA\ compared to the no Li case.}
		\label{Fig1}
\end{figure*}
		
\section{Value-Added Catalog of C3PO}
		
\subsection{Lithium abundance}
		
Our current investigation, C3PO III, investigates the relationship between planet engulfment and A(Li)\footnote{A(X) = 12 + log(N$_X$/N$_H$), where N$_X$ is the number of atoms of species X.}. Li was not previously derived in C3PO I and II due to a few subtleties, including NLTE correction, which we address in this study.
		
For the 1D-LTE Li abundance, we use the Kurucz line list\footnote{\url{http://kurucz.harvard.edu/linelists.html}}, the Kurucz stellar atmosphere models (\citealt{1992IAUS..149..225K}), and the {\it synth} task in MOOG (\citealt{1973ApJ...184..839S}) to synthesize the Li I 6707.8\AA\ doublet. To distinguish between Li detection and upper limit, we calculate the 3$\sigma$ equivalent width (EW) of Li using Equation 1 in \cite{1993ApJ...414..740D}, based on the signal-to-noise ratio near the Li line (S/N, included in Table~\ref{Table1}), full width at half maximum (FWHM), and the instrument's pixel scale:
		\begin{equation} \label{eqn1} 
			3 \sigma\ {\rm EW} =3 \times 1.503 \times \frac{\sqrt{{\rm FWHM} \times {\rm pixel}\ {\rm scale} }}{\rm S/N}
		\end{equation}
		
In instances where Li is detected, the 6707.8\AA\ absorption line is significantly stronger than the synthetic spectra with no measureable lithium feature (A(Li) = -10.0 dex). This leads to a measured equivalent width surpassing the 3$\sigma$ EW threshold, resulting in the derivation of a 1D LTE A(Li) through the best-fit synthesis. Conversely, in cases where no confident detection is shown, the Li absorption line is weak and similar to the no Li case.
		
More quantitatively, instances with measured EWs falling below the computed 3$\sigma$ EWs are categorized as non-detections, yielding a 3$\sigma$ upper limit A(Li) corresponding to this EW. Figure~\ref{Fig1} shows several examples of the syntheses of lithium detections and upper limits. For Li detections, we delineate the best-fit A(Li) (green line), A(Li) a factor of two below (orange line), and A(Li) a factor of two above the best-fit (red line). For upper limits, we show the A(Li) corresponding to the 3$\sigma$ EW (orange line). In both cases, the no Li scenario (A(Li) = -10.0 dex, cyan line) is depicted for comparison, serving as a gauge for confident detection. A more detailed procedure for deriving A(Li) and distinguishing detections from upper limits is described in \cite{2022MNRAS.513.5387S, 2023ApJ...952...71S}.
		
The uncertainty estimation in A(Li) largely follows the description in \citet{2024AJ....167..167S}. The uncertainty originates from two sources: first, uncertainty from fitting the synthesis, which can be calculated from the EW measurement uncertainty; second, uncertainty propagated from the uncertainty in the stellar parameters. The uncertainty from EW measurement at S/N $\sim$ 200 ranges approximately from 0.01 to 0.02 dex for stars with detectable A(Li). As for the uncertainty from the stellar parameters, changes in $T_{\rm eff}$ within 10-20 K correspond to $\Delta$A(Li) $\sim$ 0.01 dex, while variations of 20-30 K and 30-40 K correspond to $\Delta$A(Li) $\sim$ 0.02 dex and 0.03 dex, respectively. The precision of Li detections, co-added in quadrature, ranges from 0.01 to 0.04 dex.
		
The Li abundance measured at 6707.8\AA\ can be subject to non-local thermal equilibrium (NLTE) biases (\citealt{2016A&A...586A.156K}), we apply a correction using the publicly available BREIDABLIK package (\citealt{2021MNRAS.500.2159W}). The package takes the star's $T_{\rm eff}$, log g, [Fe/H], and 1D LTE A(Li) as inputs, and provides the 3D NLTE corrections for Li as the output. We apply the correction to our stars based on their estimated stellar parameters. The stellar parameters in our sample are within the grid computed by BREIDABLIK. The NLTE corrections typically range between -0.02 to -0.10 dex for our co-moving pairs.
		
\begin{figure*}
	\centering	\includegraphics[width=0.9\textwidth]{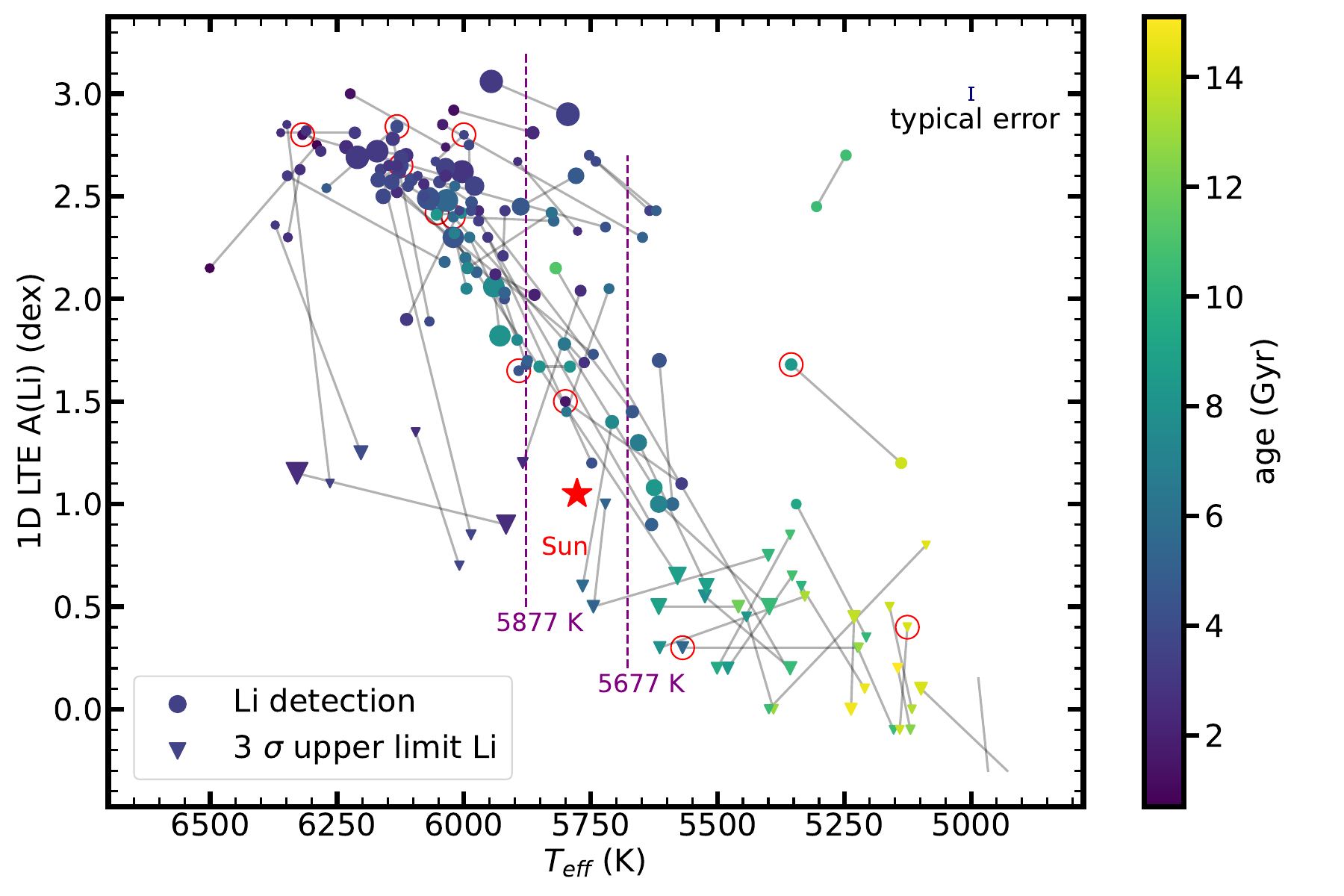}
	\caption{The 1D LTE A(Li) -- $T_{\rm eff}$ pattern for 89 co-natal pairs. Detections are marked by filled circles, while 3$\sigma$ upper limits are denoted by downward triangles. Stars are colored by their age (age1) estimation, which includes Gaia parallax and apparent magnitude, with symbol sizes proportional to {\it v} sin $i$. Each co-moving pair is linked by a gray line. Red circles highlight the companions with higher [Fe/H] in pairs displaying evidence of planet engulfment. The cooler end of the Li-Dip (6,200--6,675 K) is noticeable through several 3$\sigma$ upper limit A(Li) measurements with considerably low values. As $T_{\rm eff}$ decreases, A(Li) diminishes, eventually reaching values so low that only upper limit abundances for lithium can be determined ($T_{\rm eff}$ $<$ 5600 K). The solar $T_{\rm eff}$ (5777 K) and A(Li) (1.05 dex, \citealt{1997AJ....113.1871K}) are shown as a red star in the plot, while the $\pm$100 K range around the solar $T_{\rm eff}$ is indicated by purple lines. A(Li) shows a rapid decrease near the solar $T_{\rm eff}$. We present the 1D-LTE A(Li) as a comprehensive overview of our measurements for all co-natal pairs. While NLTE corrections cannot be applied to 3$\sigma$ upper limits, these upper limits are particularly informative for cooler and older dwarfs.}
	\label{Fig2}
\end{figure*}

Our sample boasts one of the most extensive examinations of high-precision lithium abundances in co-moving stars. In Figure~\ref{Fig2}, we show the 1D LTE A(Li) -- $T_{\rm eff}$ pattern for the 89 co-natal pairs. We note that, although the pairs of stars are co-moving, their separations are far enough ($>$ 4 $\times$ 10$^{3}$ AU) that we expect the lithium changes due to binary evolution to be minimal. Also, in C3PO I, we examined that, for individual stars, there is no obvious sign of SB2 binaries. Hence, while our co-moving pairs of stars have the advantage of being able to detect planet engulfment, the abundances can be treated individually. 
		
Figure~\ref{Fig2} shows a clear trend of decreasing Li abundance and increasing inherent scatter for late-type stars. The cooler side of the F-dwarf Li-Dip (6,200-6,675 K) is highlighted by several 3$\sigma$ upper limit A(Li) measurements. The decreasing A(Li) with $T_{\rm eff}$ pattern in our co-natal pairs conforms to findings in previous studies (e.g., \citealt{2018A&A...615A.151B, 2012ApJ...756...46R, 2023ApJ...952...71S}), which arises from the variation of Li with stellar mass (e.g. \citealt{2023MNRAS.525.4642R, 2021A&A...654A..46D, 2016A&A...590A..94C}).
		
\subsection{Rotational velocity}
		
Prior investigations have highlighted a link between projected rotational velocity ({\it v} sin {\it i}) and Li enrichment or depletion. For instance, during the main sequence (MS), angular momentum loss accompanied by differential rotation tends to deplete Li (\citealt{2023ApJ...952...71S}), while rotational mixing may contribute to its enrichment in the red giant phase (\citealt{2022MNRAS.513.5387S}). As such, to rule out any influence from rotation in our study, we also measure the {\it v} sin {\it i} value and probe the potential relationship between {\it v} sin {\it i} and A(Li), as we will elaborate below.
		
We compute {\it v} sin {\it i} by using the {\it fxcor} task in IRAF, as discussed in \cite{2020AJ....159..220S, 2022MNRAS.513.5387S}. In brief, we calculate the cross-correlation function between the object spectrum and a template in the Fourier space, fits a Gaussian to the highest peak, and determines the line broadening based on the peak’s central position and width (\citealt{1984A&A...138..183B}). The {\it v} sin {\it i} and associated uncertainty are subsequently computed from line broadening.
		
\subsection{Stellar age}
		
\begin{figure}
	\centering	\includegraphics[width=0.46\textwidth]{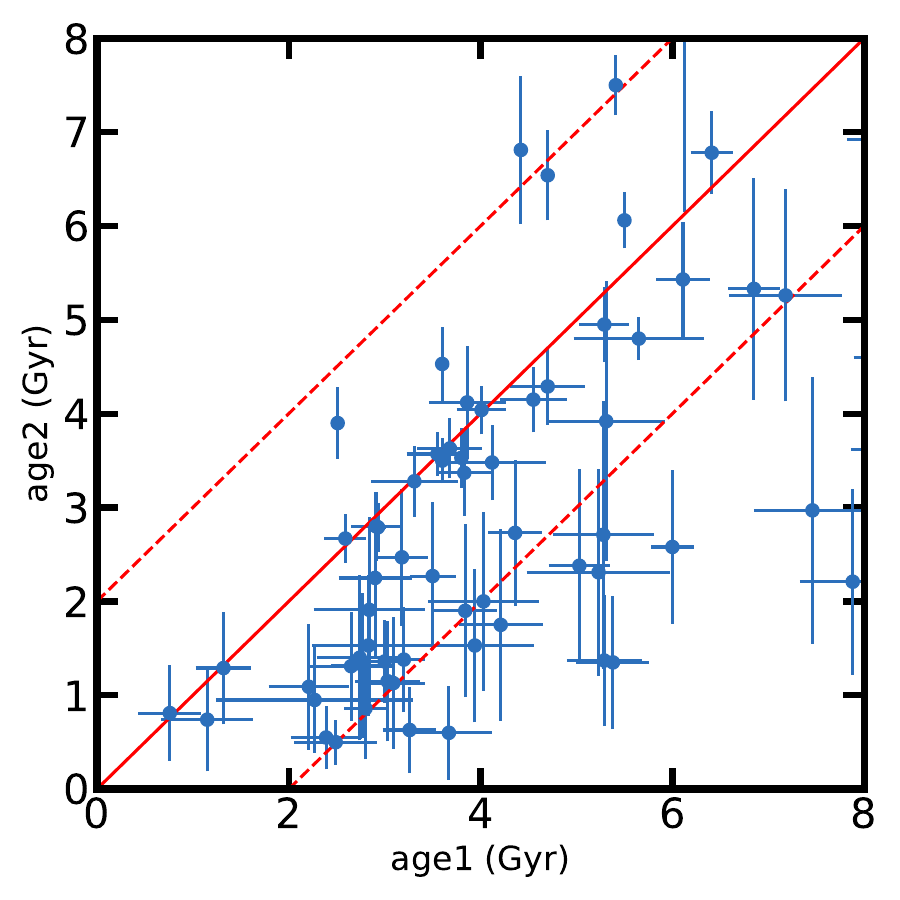}
	\caption{The comparison between age1 and age2. Age1 is calculated using absolute magnitude and $T_{\rm eff}$, with their locations on the H-R diagram compared to theoretical isochrones. Age2 is derived from spectroscopic log g and $T_{\rm eff}$, and similarly compared to theoretical isochrones on the H-R diagram. We show the lines with $|$age1 - age2$|$ $<$ 2 Gyr.}
	\label{Fig3}
\end{figure}

We adopt the q$^2$ Python routine (\citealt{Ramirez2014}) and Yonsei-Yale (Y$^2$) isochrones (\citealt{2003ApJS..144..259Y}) to derive ages. We derive two sets of ages, both listed in Table~\ref{Table1}. For the first set of ages, we use the spectroscopic $T_{\rm eff}$ and [Fe/H], pairing them with the Gaia parallax and apparent magnitude (for absolute magnitude), to compare the star's location with the stellar evolution tracks on the Hertzsprung–Russell (H–R) diagram. This method requires a precise estimate of the star's luminosity, which is affected by extinction and uncertainties in distance measurements (parallax). For the second set of ages, instead of using the Gaia data, we adopt the spectroscopically determined $\log g$, instead of the absolute magnitude, and similarly compare the star's position to the stellar evolution tracks on the H–R diagram. This method needs a precise measurement of log g, with uncertainties from the equivalent width measurements of the Fe I and Fe II lines. In addition, the range of ages (and therefore log g) near solar metallicity is wide from empirical isochrones, as pointed out by \citealt{1993ApJS...86..153G}. In both approaches, a maximum-likelihood calculation is employed to identify the most probable values for age. The methods for calculating the two sets of ages are also explained in \citet{2018MNRAS.474.2580S}. The two ages, denoted as age1, age2, and their associated uncertainties are shown in Table~\ref{Table1}. Figure \ref{Fig3} shows a comparison of the two age estimates. Generally, age1 is systematically lower than age2, although for older stars, age2 can occasionally be larger than age1. 
		
\begin{figure}
		\centering	\includegraphics[width=0.46\textwidth]{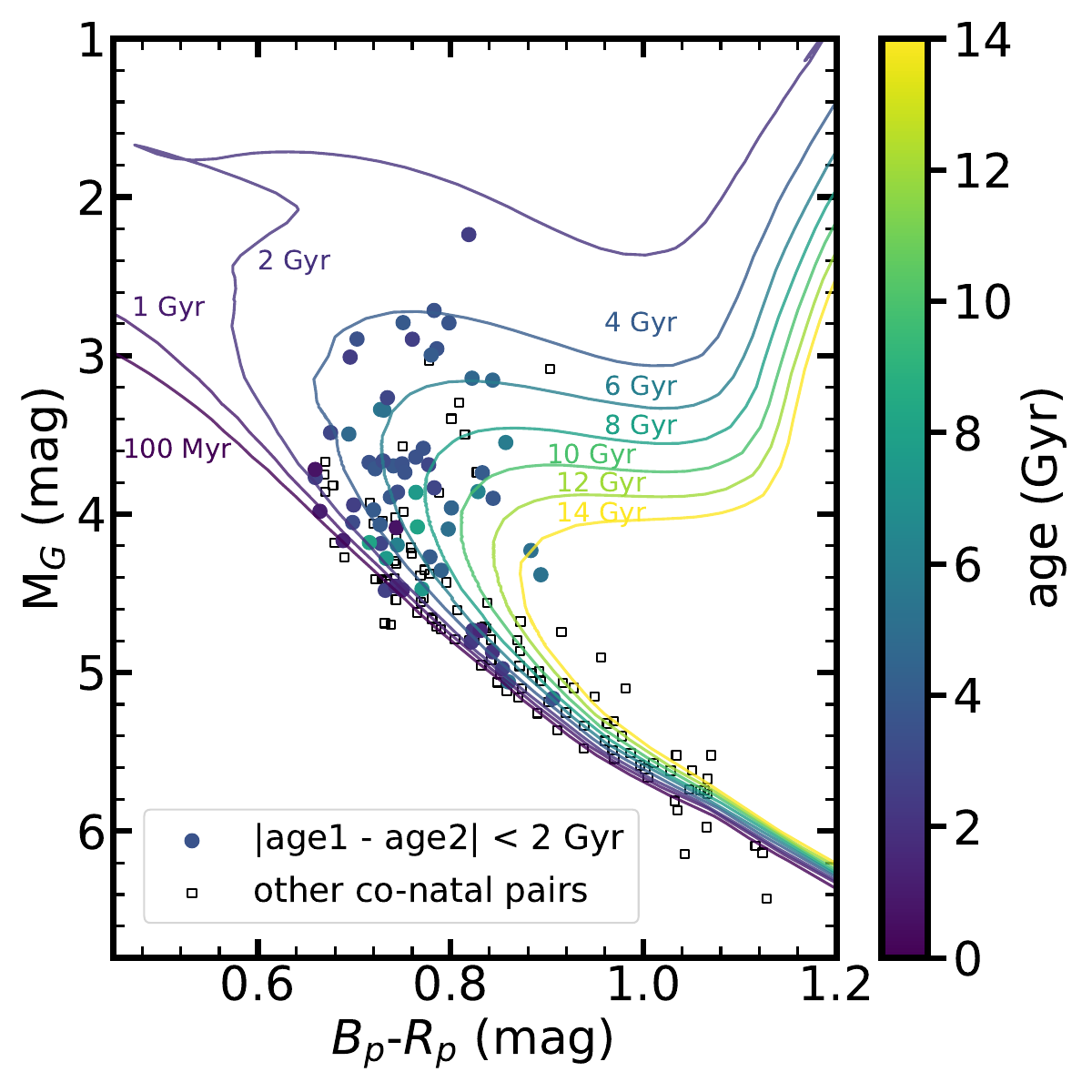}
		\caption{The Color-Magnitude Diagram for co-natal pairs, where each individual stars from the pairs are displayed. The large filled circles show stars with $|$age1 - age2$|$ $<$ 2 Gyr. They are color-coded based on their age1 values. The smaller black open squares show the rest of stars. The PARSEC isochrones \cite{2012MNRAS.427..127B} of a range of ages from 100 Myr to 14 Gyr are shown for reference.}
		\label{Fig4}
\end{figure}

In Figure \ref{Fig3}, stars positioned along the diagonal show similar measurements for age1 and age2, denoted by lines where $|$age1 - age2$|$ $<$ 2 Gyr. In Figure \ref{Fig4}, those stars are depicted as large filled circles, primarily consisting of subgiants and turnoff stars. In contrast, stars with $|$age1 - age2$|$ $>$ 2 Gyr are marked with small open squares, mainly comprising main-sequence stars, indicating considerable systematic errors in their age estimates. Additionally, several PARSEC isochrones (\citealt{2012MNRAS.427..127B}) spanning ages from 100 Myr to 14 Gyr are plotted for reference.
		
\section{Intra-pair differences in A(Li) among co-natal pairs}
		
\begin{figure*}
	\centering	\includegraphics[width=1.0\textwidth]{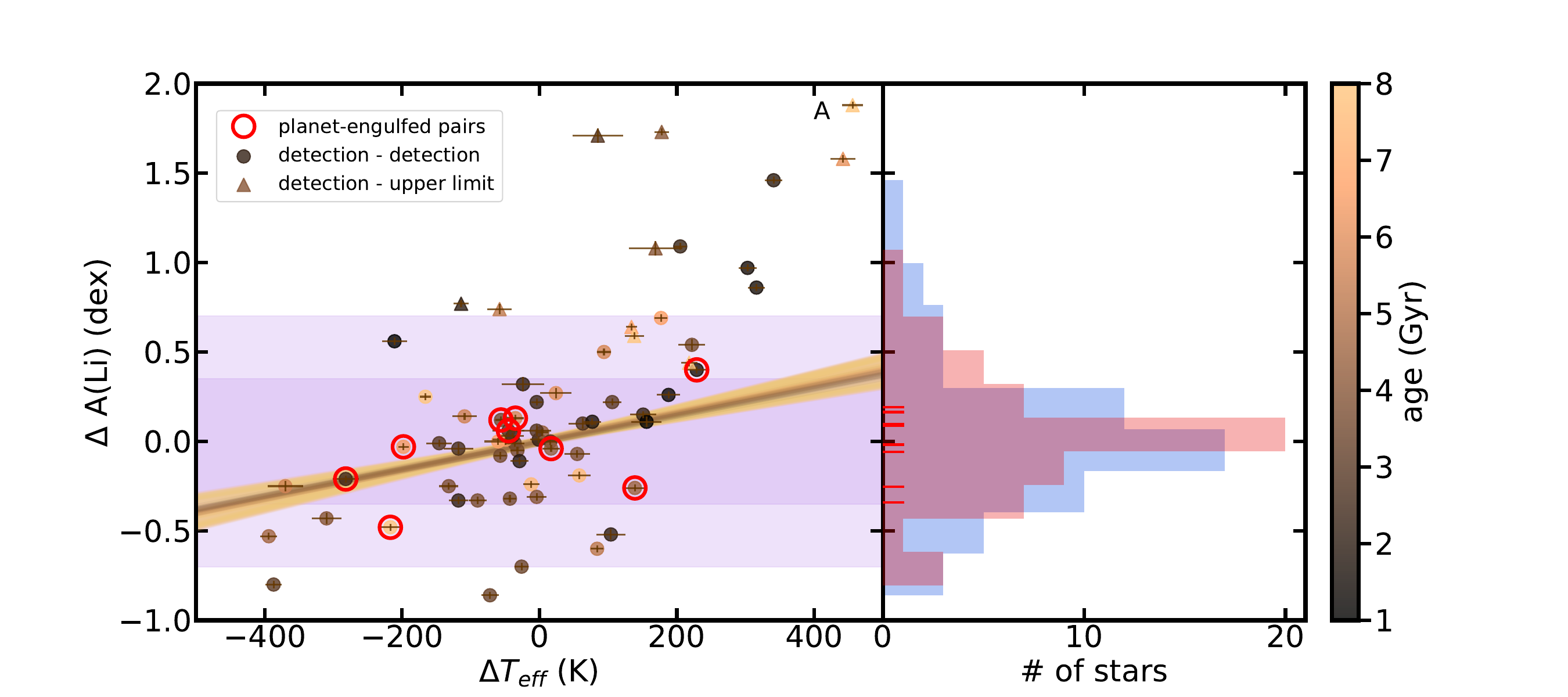}
	\caption{{\it Left panel}: Intra-pair differences in A(Li) ($\Delta$A(Li)) versus $T_{\rm eff}$ ($\Delta T_{\rm eff}$) among co-natal pairs. Filled circles show pairs with Li detections in both companions, while upward triangles denote cases with detection - upper limit, suggesting actual $\Delta$A(Li) values higher than depicted. Cases with two 3$\sigma$ upper limits for A(Li) are excluded. Pairs are color-coded based on the average age of the two companions. Pairs showing planet engulfment signatures are in red circles. Orange shaded lines show the forecasted $\Delta$A(Li) as a function of $\Delta T_{\rm eff}$ within the pair, color-coded by their stellar age. Purple-shaded bands illustrate the typical intrinsic scatter of 0.35 dex (see Figure~\ref{Fig6}) and the 2$\sigma$ level of 0.70 dex. {\it Right panel}: Marginal distribution of $\Delta$A(Li) without (blue) and with (red) $\Delta T_{\rm eff}$ influence removed according to the MDN model shown in Figure~\ref{Fig6}. For the $\Delta T_{\rm eff}$-corrected (red) distribution, we exclude the case with one A(Li) detection and one A(Li) upper limit, as our MDN model is trained solely on A(Li) detections. The seven pairs of stars with planet engulfment signatures determined in \citet{Liu2024} are plotted as horizontal ticks at the left side of the panel. After $\Delta T_{\rm eff}$ correction, most of the $\Delta$A(Li) values are largely confined within the intrinsic scatter. None of the seven pairs of stars with planet engulfment shows $\Delta$A(Li) beyond the determined intrinsic scatter, demonstrating that most planet engulfment would likely generate Li enhancement buried within the intrinsic Li scatter from other intrinsic mechanisms.}
\label{Fig5}
\end{figure*}

We focus on the intra-pair differences in NLTE-A(Li) within the 89 co-natal pairs. Among them, 59 exhibit Li detections in both companions, 11 display one Li detection and one 3$\sigma$ upper limit, while 19 pairs have Li 3$\sigma$ upper limits in both companions. The left panel of Figure~\ref{Fig5} shows the differences in A(Li) ($\Delta$ A(Li)) versus $T_{\rm eff}$ ($\Delta$ $T_{\rm eff}$) for the 70 co-natal pairs with two (shown in circles) or one (shown in upward triangles) Li detection(s), colored by their average age. Pairs with two upper limits are excluded because their locations on the plots are uncertain. 
		
This differential study between these co-natal pairs, especially those with minimal differences in $T_{\rm eff}$, demonstrates how varied Li abundances might be due to internal processes or planet engulfment. As shown in the left panel of Figure~\ref{Fig5}, most of our co-natal pairs, by selection, have $T_{\rm eff}$ differences of less than 200K, although some have larger differences. Even within those that are extremely close in $T_{\rm eff}$, there is a Li scatter of about 0.35 dex, irrespective of whether the pairs of stars have demonstrated planet engulfment signals (circled in red) or not. A similar level of Li intrinsic scatter ($\sim$ 0.35 dex) has also been observed in stars of the same $T_{\rm eff}$ and age within open clusters, such as the Pleiades (125 Myr; \citealt{2018A&A...613A..63B}) and M48 (420 Myr; \citealt{2023ApJ...952...71S}). This varied mix of detections and upper limits within the same co-natal pairs already hints at variations in chemical compositions and discrepancies in the stellar processes they have undergone.
		
The left panel of Figure~\ref{Fig5} shows a clear trend: as $\Delta T_{\rm eff}$ increases, $\Delta$ A(Li) changes accordingly. This resonates with Figure~\ref{Fig2}, where there is a clear trend of A(Li) with $T_{\rm eff}$, thus showing that at least some of the scatter is due to differences in stellar parameters, and potentially also to other auxiliary variables like {\it v} sin {\it i} and stellar ages, even when comparing $\Delta$A(Li) within co-natal pairs. Understanding the true intrinsic scatter of A(Li) thus requires us to account for such systematic effects.
		
\begin{table*}[htbp]
	\label{Table2}
	\caption{Open cluster data adopted to calibrate the Li abundance relation with respect to stellar properties}
	\begin{tabular}{cccccc}
	\hline\hline
	Name & Age & [Fe/H] & $T_{\rm eff}$ range &
				\# of Li & References \\
	& Gyr & dex & K & detection &  \\
	\hline
	Pleiades & 0.1 & 0.03 & 4200 -- 7700 & 112 & \cite{2007PhDT.........2M} \\ 
	M48 & 0.42 & -0.06 & 4350 -- 7700 & 126 & \cite{2020AJ....159..220S, 2023ApJ...952...71S} \\
	Hyades & 0.65 & 0.15 & 5380 -- 7350 & 41 & \cite{2017AJ....153..128C} \\
	Praesepe & 0.65 & 0.15 & 5320 -- 7630 & 50 & \cite{2017AJ....153..128C} \\
	NGC 3960 & 1 & 0.02 & 5200 -- 6770 & 29 & \cite{2007AA...475..539P} \\
	M67 & 4 & 0.00 & 4170 -- 6150 & 72 & \cite{2012AA...541A.150P} \\
	NGC 188 & 6.3 & 0.06 & 4961 -- 5965 & 122 & Sun et al. (2024, submitted) \\
	\hline
	\end{tabular}
\end{table*}
		
\begin{figure*}[htbp]
	\centering	\includegraphics[width=1.00\textwidth, trim={0cm 0 7.5cm 0}]{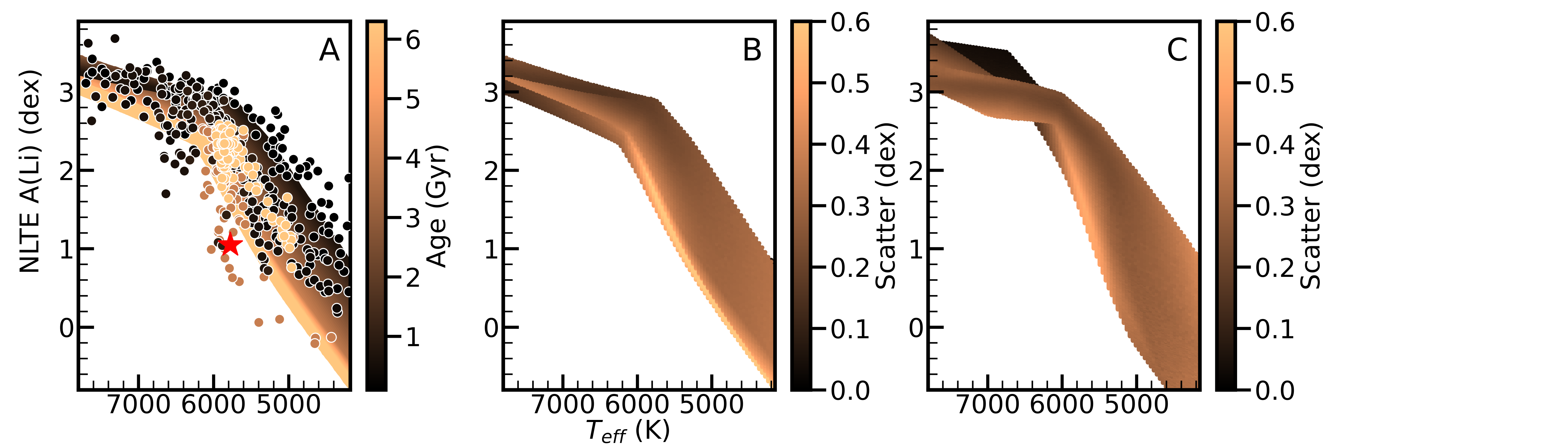}
	\caption{The probabilistic relation of lithium with respect to effective temperature and stellar age. We learn the conditional distribution of $p({\rm A(Li)} | T_{\rm eff}, \tau, v \sin i)$ and $p({\rm A(Li)} | T_{\rm eff}, \tau)$ using an MDN trained on the open cluster data (shown as solid symbols). In the left panel, the background color represents the expected mean prediction at given $T_{\rm eff}$ and $\tau$. The Sun is marked as a large red star (5777 K, 1.05 dex). The middle panel shows the scatter, defined as the $1 \sigma$ (16-84 percentile) range from the GMM at the same $T_{\rm eff}$ and age as on the left panel. On the right panel, we include the rotational velocity $v \sin i$ as an auxiliary variable, and learn the distribution $p({\rm A(Li)} | T_{\rm eff}, \tau, v \sin i)$. Plotted is the scatter of the distribution conditioned at $v \sin i =$ 10 km s$^{-1}$. The similar scatter between the middle and right panels shows that including $v \sin i$ does not reduce the intrinsic scatter, suggesting that the scatter predicated upon $T_{\rm eff}$ and $\tau$ is a robust estimate of the intrinsic scatter among co-natal stars.}
\label{Fig6}
\end{figure*}
		
\subsection{Modeling the variation of Li with stellar properties}
		
The challenge of modeling the relation and dependency of A(Li) to other measurable properties of stars lies in its fundamentally probabilistic nature. Even at a fixed set of properties, the distribution of lithium, $p({\rm A(Li)} | T_{\rm eff}, \tau, v \sin i)$ and $p({\rm A(Li)} | T_{\rm eff}, \tau)$, is not single-valued. In fact, the probability takes an unknown shape, extending beyond a simple Gaussian distribution. As such, we chose to model $p({\rm A(Li)} | T_{\rm eff}, \tau, v \sin i)$ and $p({\rm A(Li)} | T_{\rm eff}, \tau)$ with Mixture Density Networks (MDNs) (see also explanation in \citealt{Ting23}).
		
MDNs are a type of neural network that combines the flexibility of neural networks with the probabilistic output of Gaussian Mixture Models (GMMs). Unlike traditional neural networks that output point estimates, MDNs produce probability distributions as outputs. This makes them particularly suitable for our problem, where we need to capture the complex, multi-modal nature of A(Li) distributions at fixed stellar properties. The key idea of MDN here is to take the input features $T_{\rm eff}, \tau$ (and $v \sin i$) and output the parameters of a GMM (means, variances, and mixing coefficients). Essentially, one can think of MDN as a simple extension of GMM, where at fixed stellar properties, $p({\rm A(Li)} | T_{\rm eff}, \tau, v \sin i)$ and $p({\rm A(Li)} | T_{\rm eff}, \tau)$ are modeled as a Gaussian mixture model, but the properties (e.g., modes, widths) of these GMMs vary smoothly with respect to the input parameters. This approach follows the physical intuition on how the distribution of A(Li) should vary across different stellar properties.
		
Mathematically, an MDN models the conditional probability distribution as:
$$p(A(Li) | \mathbf{x}) = \sum_{k=1}^K \alpha_k(\mathbf{x}) \mathcal{N}(A(Li) | \mu_k(\mathbf{x}), \sigma_k^2(\mathbf{x}))$$
		
\noindent
where $\mathbf{x}$ represents the input features, $K$ is the number of mixture components, and here we choose $K$ to be 5. $\alpha_k$ are the mixing coefficients, and $\mu_k$ and $\sigma_k^2$ are the means and variances of the Gaussian components, respectively. All of these are functions of the input $\mathbf{x}$, where the function is parameterized by a multilayer perceptron with 20 layers, each containing 50 neurons.
		
As the stellar ages from isochrone fitting can be more noisy, especially for main sequence stars, we use data from various open clusters with reliable stellar parameter measurements, including age, across the entire $T_{\rm eff}$ range to derive the relation. These clusters are shown in Table~\ref{Table2}. The combination provides us with 552 stars spanning roughly the same temperature range as in this study. A(Li) from stars in these clusters has been corrected for the NLTE effect using the BREIDABLIK package (\citealt{2021MNRAS.500.2159W}) to match the Li abundance scale adopted in this study. To train the MDN, we adopt a learning rate of 1e-4 and a batch size of 64. We assume a 4:1 ratio of training-validation split and found that the likelihood of the validation set is the same as the training set, confirming no overfitting.
		
To check if $v \sin i$ plays a role in determining the intrinsic spread, we also fitted the MDN to study $p({\rm A(Li)} | T_{\rm eff}, \tau)$. The left panel of Figure~\ref{Fig6} shows the mean value (expectation of the distribution of $p({\rm A(Li)} | T_{\rm eff}, \tau)$) at different $T_{\rm eff}$ and $\tau$. Except at the hotter end, the MDN captures the expected behavior: cooler stars, due to their larger convective envelopes, show a larger depletion of lithium, and lithium depletes more at older ages. We caution that since there are no training data at A(Li)$\,> 3$ and $T_{\rm eff} > 7000\,$K, the mean value in this region should be viewed with caution.
		
The middle panel of Figure~\ref{Fig6} shows the measured scatter, defined as the 16-84 percentile range (or $1\sigma$) of $p({\rm A(Li)} | T_{\rm eff}, \tau)$ at the given temperature and age as shown in the left panel. We note that, since at a given temperature and age, the distribution of lithium is a multimodal Gaussian, the 16-84 percentile range is determined through sampling from the MDN. On the right-hand side, we show a similar scatter but conditioning on a $v \sin i$ value of 10 km s$^{-1}$, i.e., $p({\rm A(Li)} | T_{\rm eff}, \tau, v \sin i = 10)$. Similarly, due to the lack of hot stars, the scatter at A(Li)$> 3$ and $T_{\rm eff} > 7000\,$K collapses to a minimal value as expected, and should not be used.
		
A large part of the correlation between $v \sin i$ and lithium is due to the correlation with age, as stars slow down their rotation as they age. The probabilistic model presented here tells a similar story -- if $v \sin i$ were an important independent variable, we would expect the scatter in the right panel to be significantly smaller than in the middle panel, but this is not observed. This suggests that, at least a prominent part of the intrinsic scatter is due to $T_{\rm eff}$ and age, while the influence from other parameters is insignificant in training the model. As such, in the remainder of this paper, we will refer to the intrinsic scatter as the scatter in $p({\rm A(Li)} | T_{\rm eff}, \tau)$.
		
We acknowledge that due to the lack of older clusters, there are several limitations in this study that can only be resolved with future observations of older clusters with exquisite lithium measurements. Furthermore, since the open cluster sample is somewhat heterogeneous, there might be systematic differences between studies. However, this is an unfortunate but necessary choice, as open clusters currently provide the most accurate (and homogeneous) age estimates for main-sequence stars. Typical uncertainties for A(Li) are 0.02 -- 0.08 dex, and for $T_{\rm eff}$ are $\sim$ 50 K. While these uncertainties are relatively small, they contribute to some of the scatter observed in Figure \ref{Fig5}. However, they are subdominant, as the overall intrinsic scatter in A(Li) is $\sim$ 0.35 dex, indicating that the scatter is real.
		
Nonetheless, as shown in the middle panel, ignoring the older end, the typical intrinsic scatter in lithium even at a fixed temperature and age is of the order of 0.35 dex. This scatter is largely independent of the chosen temperature and age. Such a scatter is likely dominant compared to any heterogeneity of the open cluster sample. This scatter may explain much of the variation seen in Lithium abundances and is a key focus of this study.
		
\begin{figure*}
	\centering	\includegraphics[width=1.0\textwidth]{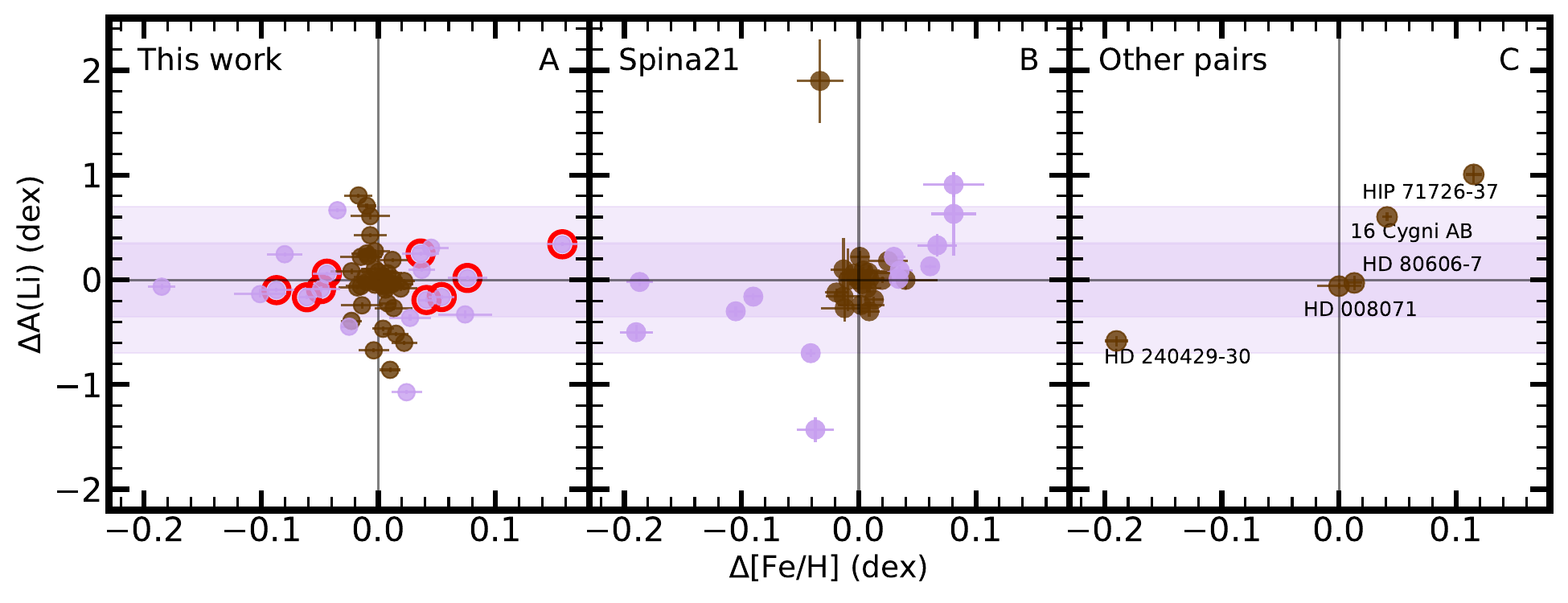}
	\caption{The intra-pair differences in A(Li) versus [Fe/H] after removing the $T_{\rm eff}$ influence on A(Li) within the pair (panel A and C). The left panel shows the sample in this study from C3PO. Chemically homogeneous pairs are denoted by brown circles, and chemically anomalous pairs by purple circles. Pairs showing evidence of planet engulfment are highlighted in red circles. Most of the sample falls within the intrinsic scatter of 0.35 dex (2$\sigma$ = 0.70 dex also shown) as probed by the MDN model (see Figure~\ref{Fig6}). As the MDN model is trained on A(Li) detections only, we exclude the several detection - upper limit cases. However, the $\Delta$A(Li) does not seem to correlate with $\Delta$[Fe/H], another indicator of planet engulfment. The pairs with planet engulfment detections are all within the range of intrinsic scatter, further confirming that planet engulfment does not generate appreciable A(Li) signals beyond the intrinsic scatter. The middle panel shows a similar study using data from \cite{2021NatAs...5.1163S}, where we do not remove the $T_{\rm eff}$ effect as $T_{\rm eff}$ of individual companions are not reported. The right panel shows a compilation of individual pairs showing evidence or potential indication of planet engulfment from the literature with measured A(Li) \citep{2019A&A...628A.126M, 2018ApJ...854..138O, 2021ApJ...922..129G, 2001A&A...377..123G, 2018A&A...614A.138L}, where the $T_{\rm eff}$ effect is removed. In contrast to this study, the literature values show some tentative $\Delta$[Fe/H] and $\Delta$A(Li) correlation, but the results might be impacted by the heterogeneous sample.}
\label{Fig7}
\end{figure*}
		
\subsection{Intrinsic Scatter vs. Planet Engulfment Signatures}
		
The probabilistic model above allows us to investigate whether the scatter seen in Figure~\ref{Fig5} can be explained by either the mean A(Li) trend with temperature and the intrinsic scatter at a fixed temperature and age, or whether pairs of stars with planet engulfment as detected in \citep{Liu2024} preferentially show more Li enhancement as proposed in other studies.
		
The orange trends show the mean A(Li) abundance changes as a function of temperature, with a range of ages from 100 Myr to 8 Gyr plotted, color-coded based on their respective ages. The purple band represents the 0.35 dex of intrinsic scatter as determined in the previous subsection. As shown, much of the A(Li) scatter, especially for pairs with $\Delta T_{\rm eff} < 100\,$K, can be largely explained by this intrinsic scatter. Pairs of stars with large $T_{\rm eff}$ differences can have larger $\Delta$A(Li), but a good part of that can be explained by the $\Delta T_{\rm eff}$ trend. There are a few outliers, none of which are those with planet engulfment signals, and these might be due to other Li enhancement mechanisms - such stellar mergers, rotational mixing, gravity waves, magnetic fields, and diffusion, which could affect lithium content in diverse ways (\citealt{2005Sci...309.2189C, 2010A&A...519L...2E, 2014ApJ...781...62L, 2019AJ....158..163D, 2021AJ....162..273S, 2022MNRAS.513.5387S, 2023A&A...672A.196K}).
		
To better demonstrate this, on the right panel, we plot the distribution of $\Delta$A(Li) of our co-natal sample estimated with a kernel density estimation using a kernel size of 0.05. The distribution in blue shows the marginal distribution without correcting $\Delta$A(Li) for the difference in $\Delta T_{\rm eff}$. Since the two stars in the co-natal pairs should share the same age, we correct them based on the MDN model with their corresponding $T_{\rm eff}$ at the average estimated age of the two stars. More precisely, we take the expectation of $p({\rm A(Li)} | T_{\rm eff}, \tau)$ at the fixed age, and the two different $T_{\rm eff}$ values lead to two different expectations (sample means) from the MDN. We then subtract the $\Delta$A(Li) based on these two different expectations, essentially correcting for the $T_{\rm eff}$ trend at the age of the pair of stars. The corrected distribution of $\Delta$A(Li) is shown in red.
		
As expected, after correcting for the difference in temperature, the distribution of $\Delta$A(Li) visibly narrows, confirming the observation from the left plot that part of the A(Li) dispersion is indeed due to temperature differences. After the correction, most of the stars fall within the 0.35 dex intrinsic scatter shown by the purple vertical band. Notably, there are some outliers in $\Delta$A(Li). While the exact cause of these outliers is difficult to determine, it's important to note that none of them appear to correlate with the planet engulfment signatures reported in \citet{Liu2024}. The $\Delta$A(Li) values of the seven pairs of stars with clear planet engulfment signatures are marked with horizontal ticks at the left of the plot, and all of these fall within the 0.35 dex intrinsic scatter.
		
A tell-tale signature of planet engulfment (a necessary but not sufficient condition) would be that pairs of co-natal stars show discrepancies in their metallicity. The correlation between the difference in metallicity, gauged by $\Delta$[Fe/H], and $\Delta$A(Li) is sometimes used as tentative proof that the difference is due to planet engulfment \citep{2021NatAs...5.1163S}. However, some of these samples, e.g., in \citep{2021NatAs...5.1163S}, are often compilations from various sources. Our vast sample of homogeneously determined [Fe/H] and A(Li), with exquisite precision ($\sigma$[Fe/H] = 0.01 dex, due to differential line-by-line analysis, and $\sigma$[A(Li) = 0.02 -- 0.03 dex), allows us to investigate this problem further. 
		
For our example, the intra-pair differences in $\Delta$A(Li) versus $\Delta$[Fe/H] within each pair are shown in the left panel of Figure~\ref{Fig7}. We have corrected $\Delta$A(Li) based on the temperature difference. The MDN model is trained only on stars with A(Li) detections, so we exclude cases that have a detection and an upper limit, as well as those with two 3$\sigma$ upper limits for A(Li). Chemically homogeneous pairs, as determined in C3PO I, are denoted by brown symbols. Chemically anomalous pairs (defined as $|\Delta$[Fe/H]$|$/$\sigma$[Fe/H] $>$ 3.0) are shown in purple symbols, while pairs showing planet engulfment signatures are indicated by red circles. Note that the uncertainties of $\Delta$A(Li) and $\Delta$[Fe/H] are negligible in this plot. 
		
Similar to the conclusion demonstrated in Figure~\ref{Fig6}, the $\Delta$A(Li) values for the engulfment pairs largely fall within the intrinsic scatter of lithium, even when temperature and age are considered. This remains true even for some extreme cases where the metallicity difference is as large as $|\Delta$[Fe/H]$|$ = 0.2 dex. While the sample of seven is too small to draw any statistical conclusions, there appears to be negligible correlation between $\Delta$A(Li) and $\Delta$[Fe/H], further suggesting that the difference in $\Delta$A(Li) is likely intrinsic due to other physical processes instead of planet engulfment. We observe some extreme outliers in $\Delta$A(Li). However, these outliers do not seem to correlate with either $\Delta$[Fe/H] or planet engulfment signatures.
		
To contrast with our results, we plot the sample from \citet{2021NatAs...5.1163S}. Their selection criteria satisfy: $\Delta T_{\rm eff} < 100$ K, $\Delta \log g < 0.1$, and $T_{\rm eff} > 6000$ K. As they do not report $T_{\rm eff}$ for individual companions in the pair (average $T_{\rm eff}$ of the pair instead), we do not correct for the difference in $\Delta T_{\rm eff}$ using the trend in our MDN model. Here we define chemically homogeneous versus inhomogeneous pairs following their definition, with $|\Delta$[Fe/H]$|$/$\sigma$[Fe/H] $\leq$ 3.0 for chemically homogeneous pairs, and values greater than 3.0 for chemically inhomogeneous pairs. Unlike our sample, there is a tentative trend between $\Delta$A(Li) and $\Delta$[Fe/H]. The exact reason for this difference is difficult to decipher, but since the sample in \citet{2021NatAs...5.1163S} is inhomogeneous, interpreting such a trend can be challenging.
		
We have also plotted in the right panel of Figure~\ref{Fig7} several studies showing evidence or potential indications of planet engulfment with $\Delta T_{\rm eff} < 200$ K, including 16 Cygni AB \citep{2019A&A...628A.126M}, HD 240429/240430 \citep{2018ApJ...854..138O}, HIP 71726/71737 \citep{2021ApJ...922..129G}, HD 008071 \citep{2001A&A...377..123G}, and HD 80606/80607 \citep{2018A&A...614A.138L}. We directly adopt NLTE A(Li) from the cited studies and apply the $\Delta T_{\rm eff}$ correction from this study. Most of these pairs show deviations in lithium that are largely consistent with our study, with the possible exception of HIP 71726-37. This system exhibits an extreme $\Delta$A(Li) of almost a dex, beyond the typical intrinsic scatter. However, such scatter, even at 1-1.5 dex, can occur within co-natal stars (e.g., the one pair at -1.1 dex in the left panel, along with the several cases with one detection and one upper limit) and is not necessarily linked to planet engulfment, as shown in our homogeneous sample.
		
\begin{figure*}
	\centering	\includegraphics[width=1.0\textwidth]{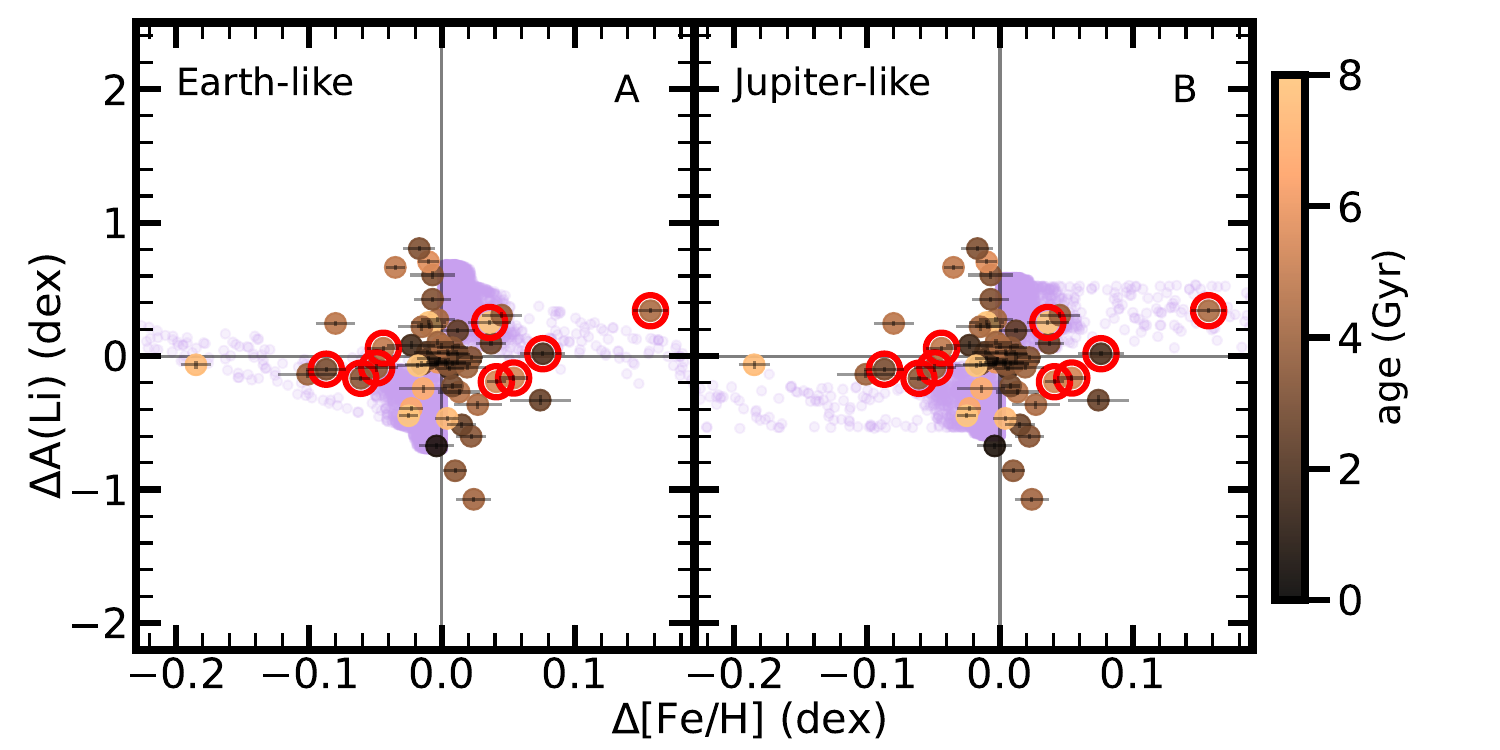}
	\caption{The expected signature of intra-pair differences in A(Li) and [Fe/H] from planet engulfment models. The lavender points show the results from simulations (see text for details) where we study the expected ejection rate of planets based on the stellar properties of the C3PO sample, and trace how the enhancement of lithium and iron from planet engulfment and their subsequent depletion might vary as a function of the star's lifetime. The left panel shows the case where we ingest Earth-like planets, and the right, Jupiter-like material. Overplotted are the same sample with $\Delta$A(Li) corrected for the $T_{\rm eff}$ trend, colored by the average of their ages (age1). Those showing planet engulfment signatures are encircled in red. The planet-engulfed pairs identified in this study closely align with the simulation results with a typical spread of less than 0.35 dex, subdominant to the intrinsic scatter of lithium, even at a fixed temperature and age, highlighting the challenge of using lithium as a tracer for planet engulfment.}
	\label{Fig8}
\end{figure*}
		
\section{Expected Li Signature from Planet Engulfment}    \label{sec5}
		
In the previous section, contrary to claims in some previous studies (\citealt{2019A&A...628A.126M, 2018ApJ...854..138O, 2021ApJ...922..129G}), our extensive homogeneous sample of co-natal stars from C3PO does not show evidence of correlation between planet engulfment and Li enhancement. In fact, from the MDN modeling, it appears that the intrinsic scatter of Li, even at the $1\sigma$ level, can be about 0.35 dex, which encapsulates much of the difference in lithium from our planet engulfment pairs. Our sample further demonstrates that in practice, some A(Li) can deviate as much as 1 dex, but without clear signs of planet engulfment as gauged by expected elemental abundance patterns from planet engulfment \citep{Liu2024}. This leads to the question: what is the expected A(Li) signature from engulfing a planet with mass consistent with that derived from the planet engulfment evidence in \citep{Liu2024}?
		
To forecast chemical abundance patterns post-ingestion of planets, especially in [Fe/H] and A(Li) following planet engulfment, we first need to assume the composition of the planets. For Earth-like planet ingestion, we adopt the composition of the Earth (\citealt{1995E&PSL.134..515A}). As for the case of ingesting a Jupiter-like planet, it is more uncertain. For simplicity, we assume the highest A(Li) value of 3.26 dex from meteorites (\citealt{2009ARA&A..47..481A}), which is close to the proto-Sun composition and adopt the same [Fe/H] as the Earth. 
		
Besides assuming the ingestion ``initial abundance," we also need to model how the enhancement gets depleted over time, as the stellar surface convection zone should gradually deplete any enhancement in the stellar atmosphere. We adopt the changes in [Fe/H] after the accretion of Earth-like planetary material using Figures 2 and 4 from \cite{2023MNRAS.518.5465B}. For lithium, we adopt the results shown in Figures 7 and 8 of \cite{2022MNRAS.516.3354S}. In both scenarios, we use the changes following the one-solar-mass case and their time evolution (due to convective depletion), and scale the signal accordingly based on the mass of the engulfed planet. Given a planet's mass and the time after accretion, the expected $\Delta$A(Li) and $\Delta$[Fe/H] can be interpolated from these models.
		
To determine the mass of planets to ingest, we adopt the planet radius distribution data from Kepler \citep{2018AJ....156..264F} and the mass-radius relationship from \citet{2019ApJ...882...38K} to establish a distribution of planet masses. We note that this mass-radius relationship, which we found to be more established, is specifically for M-type stars. There is empirical evidence showing that the mass-radius distribution of M-type stars differs from that of G-type stars (e.g., \citealt{2021AA...652A.110L}), with G-type stars producing more massive planets. Thus, our estimated planet mass distributions are likely underestimated. However, Figures 7 and 8 of \cite{2022MNRAS.516.3354S} suggest that more massive planets ($>$ 10 M$_{\oplus}$) produce similar or even smaller A(Li) overabundances than the 1 M$_{\oplus}$ case. Therefore, we contend that this planet mass distribution provides a reliable A(Li) estimate post-planet engulfment.
		
In the Kepler field, the probability of having terrestrial planets around main-sequence stars with temperatures between 4700K and 6500K, the range of temperature in C3PO, has consistently been around 50\%. Simultaneously, it has been reported that the number of planets around stars decreases with age \citep{2023AJ....166..243Y}, with the average number of terrestrial planets versus stellar age given by:
\begin{equation} \label{eqn1} 
		\overline{N}_E  = -1.58^{+0.91}_{-1.27} \times \left(\log_{10} \frac{\tau}{\rm Gyr}\right) + 3.15^{+0.79}_{-0.53}
\end{equation}
where $\tau$ is the stellar age.
		
The probability of having Jupiters around main-sequence stars in the same $T_{\rm eff}$ range is much lower. Previous studies have shown that the occurrence rate of hot Jupiters declines with age \citep{2023PNAS..12004179C}, and the average number of Jupiters as a function of stellar parameters follows:
\begin{equation} \label{eqn2}
	\begin{split}
				\overline{N}_J  = \frac{10^{5.7^{+0.4}_{-0.3}}}{10^6} \times \left(\frac{M}{M_{\odot}}\right)^{\frac{8}{3}} \times \left( \frac{R}{R_{\odot}}\right)^{-5} \\
				\times \left( \frac{P}{3\ \mathrm{day}}\right)^{\frac{13}{3}} \times \left(\frac{2.4\ \mathrm{Gyr}}{\tau}\right)
	\end{split}
\end{equation}
where $M$ and $R$ denote mass and radius of the stars, respectively, $P$ is the orbital period, and $\tau$ is the stellar age. The orbital period can be assumed to be uniformly distributed on a logarithmic scale in the [0.8, 10] day interval. For the simulation we assume planet engulfment rate = 1.
		
Given the expected number of planets as a function of time, if we assume that the planet populations at the formation stage are largely consistent over time, then to the first order approximation, the derivative over age, ${\rm d}\overline{N}/{\rm d}\tau$, represents the planet ejection rate. A subset of the ejected planets will be engulfed by the stars, with the ratio determined by the planet engulfment rate. 
		
In this study, we assume a vastly optimistic planet engulfment rate of one, i.e., all ejected planets are ingested. We note that this is certainly not a realistic assumption; however, stars that have not ingested any planets will, by definition, not have any enhancement and will only cluster at the zero location in Figure~\ref{Fig8} (assuming no intrinsic lithium scatter). Therefore, to better study the effect of planet engulfment, we chose this limit to more effectively populate the expected patterns, including some of the more extreme cases.
		
Combining all these factors to project what we should expect from the C3PO sample, we randomly select a star from the co-natal pairs and then randomly select a planet mass from the Kepler mass distribution. The stellar parameters of the selected star yield the expected planet ejection rate. From this, the planet engulfment rate determines the enhancement of $\Delta$A(Li) and $\Delta$[Fe/H] expected as a function of time. We integrate this process over time up to the selected stellar age to derive $\Delta$A(Li) and $\Delta$[Fe/H]. The time of engulfment is also chosen randomly from the star's formation to the present time, depending on its stellar age. This simulation is repeated 10,000 times for both the engulfment of Earth-like and Jupiter-like material cases separately.
		
The simulation results are shown in Figure~\ref{Fig8} as background lavender points. The left and right panels show the cases of engulfing Earth-like and Jupiter-like planets, respectively. Overplotted in Figure~\ref{Fig8} are the same data as in the left panel of Figure~\ref{Fig6} using the co-natal pairs in C3PO. Despite our grossly simplified version of the simulations, what is encouraging here is that the simulation results agree well with the pairs showing planet engulfment signatures. The lithium signal is much more subdued, at the level of 0.35 dex or lower, and therefore in many cases, is at the level of, or subdominant to, lithium intrinsic scatter, agreeing with the seven pairs of C3PO stars with claimed planet engulfment signals.
		
We note that, although in our simulation, we found a short window of time ($\mathcal{O}(10)$ Myr) during which A(Li) can significantly increase by 2-4 dex \citep{2022MNRAS.516.3354S}, which might explain some of the extreme cases in the right panel of Figure~\ref{Fig6}, lithium, unlike iron, gets depleted much more drastically. The short phase where A(Li) is greatly enhanced is statistically improbable.
		
Finally, we note that some of the outlying pairs also have the largest discrepancies in $T_{\rm eff}$. This demonstrates that despite having corrected part of the lithium-temperature trend, for large $\Delta T_{\rm eff}$, some of these corrections might still appear to be inadequate due to our relation still hinging on open clusters with a limited age range.
		
\section{Discussion}
		
In this study, we showed through the C3PO sample that stars with planet engulfment signatures do not necessarily correlate with elevated lithium abundances. In fact, much of the observed scatter appears to be explicable through intrinsic variation consistent with open cluster data. This naturally leads to two key questions: (1) What are the mechanisms that could lead to such intrinsic scatter, even at a fixed temperature and age? (2) Why are lithium abundances so subdued in planet engulfment scenarios?
		
\subsection{Mechanisms Contributing to Intrinsic Lithium Scatter}

Lithium abundances in stars depend on several factors. Li is easily destroyed in stellar interiors through proton captures at temperatures exceeding 2.5 MK. As a result, lithium only survives in the cooler surface convection zone (SCZ), making its surface abundance a valuable observational tool for studying internal stellar processes (\citealt{2017AJ....153..128C}). Li is depleted at the base of the SCZ (BSCZ), and the hotter the BSCZ , the more Li is depleted. In later-type stars (e.g., K/M dwarfs), the SCZ is deeper, leading to higher temperature at the BSCZ and more Li depletion than early-type stars (F/K dwarfs), as shown in Figure \ref{Fig2} and many previous work (e.g. \citealt{2018AA...615A.151B, 2012ApJ...756...46R}). This is the standard theory of stellar evolution (\citealt{1990ApJS...73...21D}) regarding Li.
		
The standard stellar evolutionary theory fails to explain several observed over-depletion of Li, including the very non-standard main-sequence Li-Dip (\citealt{1986ApJ...302L..49B}) and over-depletion in G/K dwarfs (\citealt{2017AJ....153..128C}), including the steep A(Li) – $T_{\rm eff}$ trend near the Sun. Beyond the standard theory, rotational mixing due to stellar spin-down (\citealt{1995ApJ...453..819R, 1997ARA&A..35..557P, 2023ApJ...952...71S}) has been proposed as the main driver of the observed lithium over-depletion. Stars rotate rapidly when they are young, when they become older they lose angular momentum and spin down, causing shear instabilities that lead to differential mixing that transports lithium from the cooler surface to the hotter BSCZ, where it is destroyed. The rate of spin-down, which varies across stars, governs the efficiency of this differential mixing and thus the rate of lithium depletion. Stars that rotate rapidly when they are young also spin down faster over time, with faster initial {\it v} sin $i$ generally leading to quicker angular momentum loss and greater lithium depletion.
		
Given this, $T_{\rm eff}$ and age are the key factors in lithium depletion. Regarding {\it v} sin $i$ the case becomes more complicated, as we do not know the star’s initial {\it v} sin $i$ and the rate of angular momentum loss over time. We can only measure the star’s current {\it v} sin $i$. Other processes, such as diffusion, gravitational settling, and stellar magnetic fields, also affect surface lithium abundances in various ways (\citealt{2014ApJ...789...53F, 2017AJ....153..128C}), further contributing to the variation seen across stars.
		
In this study, we model such variations, particularly the probability distribution of lithium abundance conditioned on temperature, age, and {\it v} sin $i$ for open cluster data. While open clusters have a limited range of ages, even in younger populations where the statistics should be more robust, our mixed density network demonstrates that the intrinsic scatter remains substantial, even when conditioning on these more easily measurable parameters. Several factors can contribute to this persistent variation:

\begin{itemize}
	\item Magnetic activity: Strong magnetic fields can suppress convection near the stellar surface (\citealt{2014ApJ...789...53F}), potentially preserving lithium from destruction. The strength and geometry of these fields can vary significantly between otherwise similar stars, leading to different lithium preservation rates. For example, in young, rapidly rotating late G/K dwarfs, the influence of magnetic fields on inflating stellar radii (\citealt{2013ApJ...765..126M}) and the effects of rapid rotation (\citealt{2019MNRAS.483.1125J, 2021MNRAS.500.1158J}) may need to be considered (\citealt{2015ApJ...807..174S, 2015MNRAS.449.4131S}).
			
	\item Early pre-main sequence accretion: Variations in accretion history during star formation can affect the initial lithium content and subsequent depletion (\citealt{2015MNRAS.454.4037T}). The chaotic nature of star formation processes can lead to differences in initial lithium abundances even among co-natal stars.
	
	\item As shown in Figure \ref{Fig4}, several stars are on the subgiant branch. Li undergoes further depletion during this phase due to the deepening of the stellar surface convection zone, as suggested by the theoretical lithium evolution model by \citet{2023A&A...677A.119D}. Furthermore, previous studies on Li and Be in the subgiant branch of M67 show that, as stars evolve into the subgiant phase, rotationally induced mixing and dilution would contribute to lithium depletion, causing variations among stellar pairs \citep{2000ApJ...544..944S, 2020ApJ...888...28B}.
			
	\item Presence of internal gravity waves: These waves can induce additional mixing (\citealt{1991ApJ...377..268G}), potentially affecting lithium abundances. The generation and propagation of these waves depend on complex internal stellar structures, which can vary among stars with similar fundamental parameters.
			
	\item Tidal effects in binary systems: Unseen close binary companions can induce tidal forces (\citealt{1995ApJ...453..819R}) that affect internal mixing and lithium depletion rates. The frequency and intensity of these interactions can vary within a single stellar population.
			
	\item Metallicity variations: Even small differences in overall metallicity (\citealt{1997ARA&A..35..557P, 2017AJ....153..128C, 2023MNRAS.522.3217M, 2024A&A...687A.234C}) can affect opacities and SCZ depths, impacting lithium depletion rate. Although co-natal stars are thought to be largely chemically homogeneous, inhomogeneities in the interstellar medium and self-enrichment from massive stars can still generate enough metallicity variation among co-natal stars to affect lithium abundances.
\end{itemize}
		
Deciphering all these factors is beyond the scope of this study. However, a detailed investigation of other observable indicators for each of these factors could provide valuable insights. For instance, magnetic field measurements through Zeeman splitting from our high-resolution spectra could help quantify the impact of magnetic activity on lithium preservation. Using asteroseismic variables as proxies for internal structure could offer clues about the presence and influence of gravity waves on lithium abundances. Additionally, high-precision astrometry could aid in identifying close binary systems, allowing us to assess the role of tidal effects on lithium depletion rates. And a much larger sample will be needed to enable more sophisticated modeling of the conditional lithium distribution with all these auxiliary variables.
		
\subsection{Lithium Generated from Planet Engulfment}
		
While it has long been known that lithium can have some intrinsic variation, perhaps an interesting discovery from our study is that, at least for the C3PO sample, stars with planet engulfment signatures do not show spectacular lithium enhancement. This enhancement appears to be limited to 0.35 dex, which is consistent with our (arguably simplistic) simulations. This suggests that lithium variation is dominated by intrinsic scatter rather than planet engulfment events.
		
While our simulations remain simplistic, they provide several insights into why this might be the case. 
		
Lithium is a fragile element that is much more easily destroyed than other elements like iron. Along with beryllium and boron, they can only survive in the outer layers of stars and serves as tracers for internal stellar processes (\citealt{2022ApJ...927..118B, 2016ApJ...830...49B}).
		
Recent studies have found that certain stars, especially those evolving into red giant phases, can produce lithium during later stages of their life cycle (\citealt{2018NatAs...2..790Y, 2022MNRAS.513.5387S}), which challenges the previous assumption that stars only destroy lithium over time. One proposed mechanism for this is planet engulfment (e.g. \citealt{2016ApJ...829..127A, 2016A&A...593A.128P}). Since planetary interiors are too cool to burn lithium, the engulfment of such a Li-rich planet could temporarily increase the star’s lithium levels (\citealt{2022MNRAS.516.3354S}). In Section \ref{sec5} and Figure \ref{Fig5}, we simulate how the star's surface lithium abundance changes after a planet is engulfed. The timescale for detecting the lithium enhancement from planet engulfment is $\mathcal{O}(10)$ Myr (\citealt{2022MNRAS.516.3354S}), during which the lithium has not yet fully mixed into the surface convective zone. As time passes the newly introduced lithium suffer from nuclear reactions at the base of the surface convection zone where temperature $>$ 2.5 MK. Additionally, the additional rotational mixing caused by planet engulfment can further accelerate the destruction of lithium, emphasizing how easily lithium is depleted within stars, especially over timescales of several billion years.
		
Our planet engulfment model is independent of whether the engulfed material is Earth-like or Jupiter-like. In the case of Earth-like material, we know that its composition has a relatively low lithium abundance, similar to that of the Sun (\citealt{1997AJ....113.1871K}). For Jupiter-like material, although the meteoritic composition suggests a higher lithium abundance (A(Li) = 3.26 dex, \citealt{2009ARA&A..47..481A}), the low hydrogen content and higher heavier elements content leads to a low lithium mass fraction. As a result, for both the Earth-like and Jupiter-like material, the lithium enhancement would be difficult to detect. In both cases, the enhancement is short-lived, and the lithium is eventually destroyed at the base of the stellar surface convection zone throughout time.
		
\subsection{Caveat and limitation}
		
While this study presents a unique and state-of-the-art sample of co-natal stars with high-quality abundance measurements, our results are predicated on several key assumptions. Notably, we have assumed that the detection of planet engulfment signatures in C3PO II \citep{Liu2024} are indeed due to planet engulfment events. In the original paper, the authors argued that these detections are consistent with the ingestion of planetary material, with Bayesian evidence significantly favoring this hypothesis over both the null hypothesis (attributing the signal to noise) and atomic diffusion. Nonetheless, it remains possible that such ingestion of planetary material might have occurred during the planet formation process (e.g., due to depletion of refractory elements accreted onto the host stars, \citealt{2023A&A...676A..87H}), rather than through late-time planet engulfment. In this scenario, the star's evolution would follow a more typical track, which could explain why the observed lithium scatter aligns with the expected intrinsic variation.
		
However, as also noted in C3PO II, the planet engulfment hypothesis may be more favorable. Our simulations demonstrate that a late-time accretion event would be necessary to create the observed scatter in heavy elements like iron (0.05-0.1 dex). It would be challenging to account for such deviations if the signal were due to early-stage planet formation processes, given the gradual depletion of elements over the stars' lifetimes. Importantly, even if the majority of the claimed planet engulfment signatures were instead attributable to planet formation processes, the lithium signature would remain subdued and largely unobservable (\citealt{2023A&A...676A..87H}). In this case, the conclusions of our study regarding the challenges of using lithium as a reliable indicator of planet engulfment would still hold.
		
Aside from the planet engulfment candidates, a notable number of C3PO star pairs show significant differences in lithium, with $\Delta$A(Li) above one dex. This remains true even after correcting for the temperature effect using our MDN model. We found that 7 out of the 8 pairs with such deviations also have the largest temperature differences (as can be seen in the left panel of Figure~\ref{Fig5}). This scatter is larger than the expected intrinsic variation, and there are multiple potential explanations for this observation.
		
One possibility is that our MDN model, which was necessarily trained only on open cluster data, performs well within the parameter range spanned by these clusters. However, as the age range is limited and quantized, this might lead to insufficient correction for some stars. We emphasize that our results, as seen in Figures~\ref{Fig5} and~\ref{Fig8}, remain consistent if we simply apply a $\Delta T_{\rm eff}$ cut of, say, 100K to the C3PO sample. While this would limit the number of samples in this study, our conclusions remain unaltered.
		
Another possibility is that the NLTE correction is inadequate. Lithium abundances can be affected by NLTE effect to varying degrees (\citealt{2020A&A...638A..58M, 2018A&A...618A..16H}), depending on different stellar atmosphere. In this study, we used a publicly available NLTE correction for late-type stars in GALAH DR3 (\citealt{2021MNRAS.500.2159W}), taking into account the different corrections for various stellar parameters. However, it is possible that incorrect corrections could create this differential difference in Li as a function of $\Delta T_{\rm eff}$. As we continue to expand the C3PO sample, we might resolve this issue by limiting our analysis to clear co-natal stellar twins in the future.

\section{Conclusion}
		
The lithium abundance in stars can be affected by myriad internal processes, as well as potential planet engulfment events. Regarding the latter, the correlation between deviations in metallicity and corresponding deviations in Li is often treated as a signature of planet engulfment. However, as lithium can also be influenced by other factors, understanding how much of the lithium variation is due to internal processes or binary evolution, and how much can be attributed to planet engulfment, remains unresolved.
		
The challenge here lies in modeling internal processes and determining the amount of lithium enhancement. This requires reference stars with roughly the same stellar properties, including effective temperature and age. However, until now, there has been a lack of detailed and systematic surveys with exquisite spectra at high resolution and high signal-to-noise ratios conducted on co-natal stars. Previous studies on the impact of internal stellar evolution on lithium abundance have been limited to open clusters.
		
In this study, we leverage the recent sample from the C3PO project, which has conducted a near-complete magnitude-limited survey of comoving stars. This project has greatly enhanced the sample of observable co-natal stars in the solar neighborhood using 6-10m telescopes \citep{2023MNRAS.526.2181Y}. Simultaneously, the differential study of these pairs has allowed for the detection of stars that have undergone planet engulfment through their distinctive elemental abundance patterns \citep{Liu2024}.
		
We derived detailed lithium abundances for 125 pairs (250 stars) in C3PO with a precision of 0.01-0.04 dex, 89 pairs of which are determined to be co-natal. Combined with the previous detection of seven pairs of stars with planet engulfment signals, this provides us with the first extensive and homogeneous sample where we can differentially study the Li deficit or enhancement between pairs of stars along with their other properties, allowing us to investigate the various factors that could lead to lithium abundance variations.
		
Our key findings can be summarized as follows:
\begin{itemize}
	\item Among the co-natal stars, there is a scatter largely confined to $\pm 0.5$ dex, with a few outliers. However, a significant portion of this scatter can be attributed to the temperature differences between the stars.
			
	\item Taking into account the temperature influence, the scatter reduces to about $\pm 0.35$ dex, which is consistent with the scatter modeled using a mixed density network for open cluster members. 
			
	\item Even at a given fixed temperature and age, the mixed density network reveals an intrinsic scatter of $\pm 0.35$ dex in lithium that can be attributed to a myriad of internal processes alone, which is consistent with our observations.
			
	\item We do not observe any correlation between the differences in $\Delta$A(Li) and $\Delta$[Fe/H], supporting the idea that the scatter in lithium is unrelated to planet engulfment alone, as we would expect a correlation between these abundances if it were.
			
	\item All seven pairs of stars with planet engulfment signatures also show lithium differences that are confined within the 0.35 dex of expected intrinsic scatter, and they also show little correlation between $\Delta$A(Li) and $\Delta$[Fe/H], at odd with some of the previous claims.
\end{itemize}
		
Further, we also estimate the expected lithium and iron enhancement from planet engulfment, taking into account the latest estimates of planet ejection rates, as well as the lithium and iron depletion rates due to stellar atmospheric convection. Our model, despite being crude, agrees with the data and suggests that most planet engulfment events, while demonstrating stronger enhancement in iron and heavier elements, show lithium enhancement confined to about 0.35 dex, at the level of, or subdominant to, the intrinsic scatter in lithium.
		
This study highlights the challenge and difficulty of relying on lithium to justify the detection of planet engulfment. That said, we do not refute that some stars, especially those caught soon after the engulfment phase, could demonstrate significantly enhanced lithium abundances. The simulations also show that this enhancement can last for $\mathcal{O}(10)$ Myr. Certainly, part of the lithium variation can be attributed to planet engulfment. However, the key message of our investigation shows that such interpretation of lithium variation should be treated with caution, as it is likely to be of the same magnitude as, or less than, the intrinsic scatter.
		
Future studies enhancing the sample with detailed ages, such as through asteroseismic samples, and investigating lithium's relation with other observables that can be tell-tale gauges of some of these intrinsic variations, would be critical to decipher this mysterious yet story-rich fragile element in the Universe.
				
\section*{acknowledgements}
		
We appreciate helpful comments from Bertram Bitsch on disk accretion. Q.S. is supported by the National Key R\&D Program of China, No. 2024YFA1611800. Y.S.T is supported by the National Science Foundation under Grant No. 2406729. F.L. and A.I.K. acknowledge support from the Australian Research Council Centre of Excellence for All Sky Astrophysics in 3 Dimensions (ASTRO 3D), through project number CE170100013. This work has made use of data from the European Space Agency (ESA) mission Gaia (\url{https://www.cosmos.esa.int/gaia}), processed by the Gaia Data Processing and Analysis Consortium (DPAC, \url{https://www.cosmos.esa.int/web/gaia/dpac/consortium}). Funding for the DPAC has been provided by national institutions, in particular, the institutions participating in the Gaia Multilateral Agreement. This paper includes data gathered with the 6.5-m Magellan Telescopes located at Las Campanas Observatory, Chile, kindly supported by Carnegie Observatories. Some of the data presented herein were obtained at the W. M. Keck Observatory, which is operated as a scientific partnership among the California Institute of Technology, the University of California, and the National Aeronautics and Space Administration. The Observatory was made possible by the generous financial support of the W. M. Keck Foundation. Based on observations collected at the European Southern Observatory under ESO programme 108.22EC.001.
		
\bibliography{sun24_planet}{}
\bibliographystyle{aasjournal}
		
\end{document}